\documentclass[aps,prl,twocolumn,nopacs,floatfix]{revtex4-2}
\usepackage{epsfig}
\usepackage{graphicx}
\usepackage{dcolumn}
\usepackage{amsthm,amsmath,amsfonts}
\usepackage{mathtools}
\usepackage{comment}
\usepackage[colorlinks]{hyperref}
\usepackage{xcolor}
\usepackage{orcidlink}
\usepackage{rotating}
\usepackage{cancel}
\usepackage[normalem]{ulem}

\begin{document}

\title{Comprehensive Assessment of $\mathrm{Th}^{3+}$ Properties for Nuclear Clock and Fundamental Physics Applications}

\author{A. Chakraborty \orcidlink{0000-0001-6255-4584}}
\email{arupc794@gmail.com}

\author{B. K. Sahoo \orcidlink{0000-0003-4397-7965}}
\email{bijaya@prl.res.in}

\affiliation{Atomic, Molecular and Optical Physics Division, Physical Research Laboratory, Navrangpura, Ahmedabad 380009, India}

\begin{abstract}
By employing singles, doubles, and triples excitations within the relativistic coupled-cluster framework, we perform comprehensive calculations of a wide range of atomic properties for the Th$^{3+}$ ion. These properties are essential for advancing nuclear clock technology and probing fundamental physics. Combining our isotope shift parameters with experimental data, we estimate highly accurate values of the differential nuclear charge radii for $^{232,229}$Th and $^{229m,229}$Th. Additionally, we determine the nuclear magnetic dipole and electric quadrupole moments for both the ground and isomeric states of $^{229}$Th by combining measured hyperfine structure constants with our theoretical calculations. Our precise evaluations of electric dipole polarizabilities and hyperfine-induced quadrupole moments are critical for assessing systematic uncertainties in $^{229}$Th$^{3+}$-based nuclear clock. Notably, we observe unexpectedly significant contributions from higher-order relativistic effects and excitations involving orbitals with higher angular momentum, which markedly influence the energies of the ground state and its fine-structure partner. These results highlight the substantial challenges in achieving highly accurate predictions for these properties.
\end{abstract}

\date{\today}

\maketitle 

\begin{figure}[t]
\centering
\includegraphics[width=8cm, height=6cm]{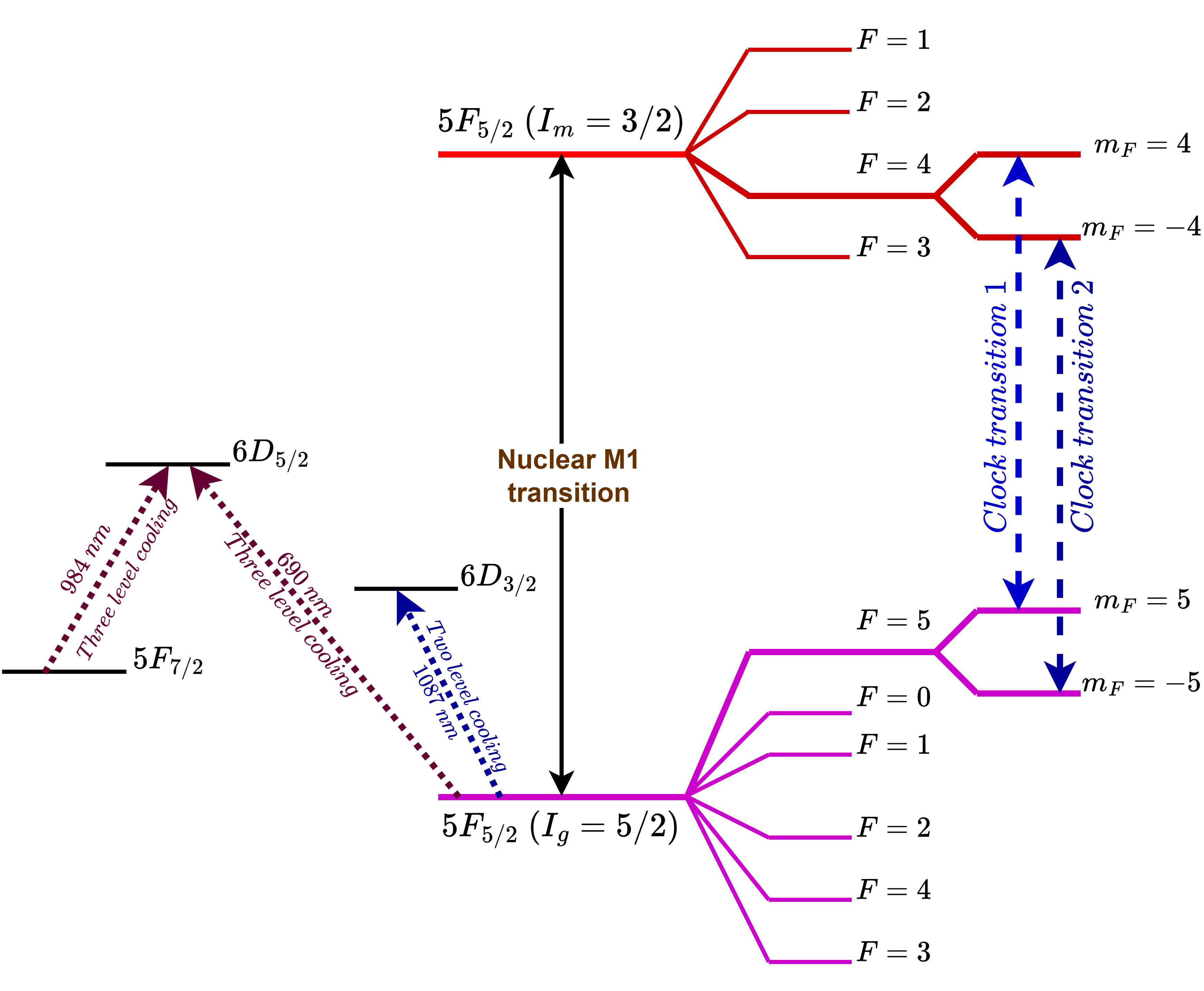}
\caption{Energy level diagram of the $^{229}$Th$^{3+}$ ion with the proposed clock and cooling transition lines involving the ground and isomeric 5$F_{5/2}$ manifolds.}
\label{fig-clock}
\end{figure}

The thorium isotope $^{229}\mathrm{Th}$ possesses a unique low-energy isomeric state at approximately 8.3 eV \cite{Kraemer2023, Peik2003, Peik2021}, making it the only known candidate for an optical nuclear clock based on this transition. The proposed clock scheme exploits the transition between the states $|5F_{5/2}, I_{g}=5/2, F=5, m_F=\pm5\rangle \rightarrow |5F_{5/2}, I_{m}=3/2, F=4, m_F=\pm4\rangle$ in triply ionized $^{229}$Th$^{3+}$, involving the nuclear ground and isomeric states with spins $I_g$ and $I_m$, respectively (see Fig. \ref{fig-clock}). This clock could reach fractional uncertainties as low as 
$10^{-19}$ \cite{Campbell2012} and offers exceptional sensitivity to potential temporal variations of fundamental constants, as well as to dark matter interactions \cite{Flambaum2006, Safronova2018}. The francium-like electronic configuration of 
$^{229}\mathrm{Th}^{3+}$ enables accurate \textit{ab initio} calculations, while its atomic energy levels are well-suited for laser cooling and trapping, establishing an ideal platform for high-precision spectroscopy \cite{Campbell2011, Zitzer2025, Safronova2013}. Experimentally, two primary methods are considered for cooling $^{229}\mathrm{Th}^{3+}$ ions for spectroscopic studies \cite{Peik2003, Zitzer2025}: (i) a closed two-level transition 
$5F_{5/2} \rightarrow 6D_{3/2}$, accessible with a 1087 nm laser, and (ii) a closed three-level ($\Lambda$-type) transition $5F_{5/2} \rightarrow 6D_{5/2} \rightarrow 5F_{7/2}$,  requiring lasers at 690 nm and 984 nm wavelengths.

From a fundamental physics perspective, precise characterization of both nuclear and electronic properties of thorium isotopes is essential. In particular, accurate determination of isotope shift (IS) parameters plays a crucial role, as these parameters provide vital insights into nuclear structure and electron-nucleus interactions within thorium isotopes. Despite their significance, relatively few studies have focused on calculating the IS constants of Th$^{3+}$; most existing research has predominantly addressed field shift (FS) constants, with limited attention paid to mass shift (MS) constants \cite{Berengut2009, Dzuba2023, Si2025}. Notably, there is an approximately 8\% discrepancy between reported values of the FS constant ($F$) for the electronic ground state \cite{Dzuba2023, Si2025}, and about a 7\% difference in the estimated change in the root-mean-square (rms) nuclear charge radius, $\delta\langle r^2 \rangle^{229m,229}$, between the isomeric and ground states of $^{229m,229}$Th, as reported in Refs. \cite{Dzuba2023} and \cite{Safronova2018-1}. This highlights the urgent need for more accurate and consistent evaluations of IS constants in Th$^{3+}$.

The leading-order IS in a transition in two isotopes $A$ and $A'$ is given by \cite{King,Sahoo-review}
\begin{equation}\label{linear}
\delta \nu^{AA'} \approx  F \, \delta \langle r^2 \rangle^{AA'}  + K^{MS} \, \mu^{AA'} ,
\end{equation}
where  $\mu^{AA'}=\frac{1}{M_{A'}}-\frac{1}{M_A}$ with $M_{A/A'}$ are masses of isotopes $A/A'$ and $K^{MS}$ is the MS constant, which is decomposed into the normal mass shift (NMS) and the specific mass shift (SMS) components: $K^{MS}=K^{NMS}+K^{SMS}$.

\begin{table}[t]
\setlength{\tabcolsep}{1pt}
\centering
\caption{Energies and IS constants of the ground and low-lying excited states of Th$^{3+}$ at different levels of approximation. The energy values are compared with experimental data \cite{NIST, Blaise} to assess their accuracy, while the FS constants are contrasted with existing theoretical calculations.}
\begin{tabular}{lrrrr}
\hline\hline
Method & 5$F_{5/2}$ & 5$F_{7/2}$ & 6$D_{3/2}$ & 6$D_{5/2}$  \\[0.3ex]
\hline
 \multicolumn{5}{c}{Energy (in cm$^{-1}$)} \\[0.5ex]
DHF  & $-206642$ & $-203213$ & $-211800$ & $-207575$ \\
RCCSD & $-227266$ & $-222883$ & $-221669$ & $-216327$ \\
RCCSDT & $-227578$ & $-223257$ & $-221590$ & $-216335$ \\[0.3ex]
+Basis & $-1201$ & $-1197$ & $-362$ & $-323$ \\[1ex]
+Breit & $-714$ & $-847$  & $-44$ & $-122$ \\[0.3ex]
+VP & $38$ & $37$ & $16$ & $14$ \\[0.3ex]
Final & $-229455$ & $-225264$ & $221980$ & $216766$  \\[0.3ex]
Experiment & $-231065$ & $-226740$ & $-221872$ & $-216579$ \\[0.5ex]
Accuracy & 1\%  & 1\% & 2\% & 2\% \\
\hline
 \multicolumn{5}{c}{$F$ (in GHz/fm$^2$)}   \\[0.5ex]
DHF & 51.26 & 49.06 & 21.90 & 19.16 \\
RCCSD & 53.61 & 51.47 & 21.5 & 19.60 \\
RCCSDT & 53.09 & 50.96 & 21.34 & 19.48  \\[0.3ex]
+Basis & 0.25 & 0.17 & 0.09 & 0.12 \\[1ex]
+Breit & $\sim$ 0.0 & $\sim$ 0.0 & 0.08 & $-0.15$  \\[0.3ex]
+VP & 0.77 & 0.60 & 0.25 & 0.09 \\[0.3ex]
Final & 54.1(15) & 51.7(12) & 21.8(10) & 19.5(10)\\[0.3ex]
SD$+$CI \cite{Dzuba2023} & 55.0(55) & 53.0(53) & 23.3(23) & 20.5(20) \\[0.3ex]
CI \cite{Si2025} & 51.4 &   &  &\\[0.3ex]
\hline
\multicolumn{5}{c}{$K^{NMS}$ (in GHz-amu)}  \\[0.5ex]
DHF & 3609.8 & 3624.1 & 3500.2 & 3456.4 \\
RCCSD & 3926.4 & 3933.4 & 3637.1 & 3580.4 \\
RCCSDT & 3937.7 & 3945.3  &  3640.5 & 3584.4 \\[0.3ex]
+Basis & $ -6.1 $ & $-4.3 $ & 4.3 & 13.3 \\[1ex]
+Breit &  35.0 & 41.5 & 1.4 & 5.6 \\[0.3ex]
+VP & $-3.8$ & $-3.6$ & $-1.8$ & $-5.2 $\\[0.3ex]
Final & 3963(84) & 3979(75) & 3644(78) & 3598(70) \\[0.3ex]
\hline
 \multicolumn{5}{c}{$K^{SMS}$ (in GHz-amu)} \\[0.5ex]
DHF  & $-9367.7$ & $-9074.5$ & $-3097.2$ & $-3169.4$ \\
RCCSD & $-8743.8$ & $-8549.4$ & $-2621.4$ & $-2727.6$  \\
RCCSDT & $-8852.6$ & $-8669.4$ & $-2640.0$ & $-2739.9$ \\[0.3ex]
+Basis & 19.1 & 15.2 & $-5.1$ & $-5.2$  \\ [1ex]
+Breit & $-121.8$ & $ -109.5$ & $ -13.7$ & $-12.1$ \\[0.3ex]
+VP & 0.9 & 0.7 & $-0.5$ & $-0.5$ \\[0.3ex]
Final & $-8954(165)$ & $-8763(150)$ & $-2659(100)$ & $-2758(109)$ \\[0.3ex]
\hline\hline
\end{tabular}
\label{tab1}
\end{table}

By incorporating the relativistic forms of the MS operators and the Fermi-charge distribution for the FS operator (see Ref. \cite{Sahoo-review} for the explicit expressions), the IS constants are estimated using the finite-field (FF) approach through an expansion
\begin{eqnarray}
E_v(\lambda) = E_v(0) + \lambda \langle O^{IS} \rangle+  {\cal O}(\lambda^2)  ,
\label{eqn1}
\end{eqnarray}
where $E_v(0)$ is the energy of an atomic state with valence orbital $v$ due to the atomic Hamiltonian $H_{at}$, $E_v(\lambda)$ is the energy obtained from the total atomic Hamiltonian $H_{\lambda} = H_{at} + \lambda O^{IS}$ with $\lambda$ is an arbitrary parameter, and $O^{IS}$ represents the corresponding IS operator. In this work, the atomic Hamiltonian $H_{at}$ is initially considered to be the Dirac-Coulomb (DC) Hamiltonian and correction terms from the Breit and vacuum polarization (VP) interactions are subsequently incorporated. Calculations are first performed at the singles and doubles (RCCSD) approximation within the relativistic coupled-cluster theory, followed by more comprehensive calculations including singles, doubles, and triples (RCCSDT method) starting from the Dirac-Hartree-Fock (DHF) method. Differences between the RCCSD and RCCSDT methods would highlight the significance of triple excitations in enhancing the accuracy of the results. A detailed description of these methods can be found in Refs. \cite{Sahoo-review,Chakraborty2025-Li,Chakraborty2025-Cs,Katyal2025}. Due to computational constraints, orbital excitations are usually restricted to orbitals with lower orbital angular momentum ($l$) values. However, it is later demonstrated that orbitals with $l>6$ contribute significantly to the energy calculations of Th$^{3+}$. To illustrate this, initial calculations are performed using orbitals up to $l=6$, followed by calculations including orbitals up to $l=9$. The differences between these results are referred to as the “$+$Basis” contributions.

\begin{table}[t]
\setlength{\tabcolsep}{8pt}
\centering
\caption{The weighted $\delta\langle r^2\rangle^{232,229}$ value inferred by combining our IS factors with IS measurements from Ref. \cite{Campbell2011}.}
\begin{tabular}{crrr}
\hline\hline
Transition & IS \cite{Campbell2011}& \multicolumn{2}{c}{This work (in fm$^2$)}\\
\cline{3-4}
&&&\\
& (in MHz) & $\delta\langle r^2\rangle^{232,229}$ & $\overline{\delta\langle r^2\rangle}^{232,229}$\\
\hline
5F$_{5/2}\leftrightarrow$6D$_{3/2}$& $-9856(10)$ & 0.316(19)  &\\[0.3ex]
5F$_{5/2}\leftrightarrow$6D$_{5/2}$& $-10509(7)$ &  0.313(16) & 0.316(9) \\[0.3ex]
5F$_{7/2}\leftrightarrow$6D$_{5/2}$& $-9890(9)$ & 0.317(16)  &\\[0.3ex]
\hline\hline
\end{tabular}
\label{tab-dr}
\end{table}

Our calculated energies for various states of Th$^{3+}$, obtained at different level of approximations, are summarized in Table \ref{tab1}. Notably, the energy of the ground $5F_{5/2}$ state is estimated to be higher than that of the excited $6D$ states at the DHF method; however, accurate trend is achieved after including correlation effects. As previously indicated, unusually large contributions from $+$Basis to the energies are observed, consistent with the findings reported in Ref. \cite{Porsev2021}. Breit interaction contributions to the energies of the ground state and its fine-structure partner are found to be significant, whereas VP contributions are moderate. A comparison between our final results and the experimental values from Refs. \cite{NIST, Blaise} shows that our calculated energies agree within 1–2\%, which could potentially be improved by incorporating higher-level excitations in our RCC calculations.

\begin{table}[t]
\setlength{\tabcolsep}{2pt}
\centering
\caption{Calculated $A_{hf}/\mu_I$ values (in MHz/$\mu_N$) for both the ground and excited states of  $^{229}$Th$^{3+}$ using the DHF, RCCSD, and RCCSDT methods. Corrections due to $+$Basis, Breit, VP, and BW are also included. Our final results are compared with previous calculations. By combining our $A_{hf}/\mu_I$  values with the measured $A_{hf}$ values, we extract $\mu_I$ of $^{229}$Th. The weighted average $\mu_I$ is then recommended and compared with values reported in the literature.}
\begin{tabular}{lrrrr}
\hline\hline
Method & 5$F_{5/2}$ & 5$F_{7/2}$ & 6$D_{3/2}$ & 6$D_{5/2}$ \\[0.3ex]
\hline
DHF & 202.96 & 105.16 & 332.44 & 121.80\\
RCCSD & 225.44 & 80.96 & 430.12 & $-30.80$ \\
RCCSDT & 224.11 & 80.89 & 432.70 & $-36.22$ \\[0.3ex]
+Basis & $-0.56$ & $0.34$ & 1.04 & 0.68 \\[1ex]
+Breit &  3.56 & 2.28 & 4.00 & 1.36\\[0.3ex]
+VP &  0.12 & $-0.24$ & 0.80 & $-1.04$\\[0.3ex]
+BW &  $-0.72$ & 0.92 & $-4.36$ & 5.24 \\[0.3ex]
\hline
Final & 226.5(4) & 84.2(8) & 434.2(28)& $-30.0(67)$ \\[0.5ex]
Ref. \cite{Safronova2013} & 229.2 & 86.1 & 431.5 & $-36.7$ \\[0.3ex]
Ref. \cite{Li2021} & 230.53 & 87.12 & 431.85 & $-23.31$ \\[0.3ex]
Ref. \cite{Porsev2021}  & 224.8(1) & 83.6(12) & 425.2(48) & $-41.6(88)$ \\ \hline \\ 
$A_{hf}$ (Expt) \cite{Campbell2011} & 82.2(6) & 31.4(7) & 155.3(12) & $-12.6(7)$ \\[0.3ex]
$A_{hf}$ (Expt) \cite{Zitzer2025} & 82.0(2) &   &   & $-12.9(3)$ \\ \hline \\
Extracted $\mu_I$ &  0.3620(11) & 0.3729(90) & 0.3576(36) & 0.430(96) \\
\hline 
\multicolumn{5}{c}{Recommended $\mu_I$ value in $\mu_N$} \\
This work &\multicolumn{4}{c}{0.3618(12)} \\ [0.3ex]
Others &\multicolumn{4}{c}{0.360(7) \cite{Safronova2013}, 0.359(9) \cite{Li2021}, 0.3663(60) \cite{Porsev2021}  } \\ [0.3ex]
&\multicolumn{4}{c}{0.54 \cite{Chasman1977}, 0.45(4) \cite{Gerstenkorn1974}} \\ [0.3ex]
\hline\hline
\end{tabular}
\label{tab2}
\end{table}

In Table \ref{tab1}, we also present our calculated $F$ constants, as well as $K^{\mathrm{NMS}}$ and $K^{\mathrm{SMS}}$ values for the investigated states of Th$^{3+}$ at the same level of approximation used for the energy calculations. Notably, for the FS, the VP contributions exceed that from the Breit interaction and $+$Basis. We compare our computed $F$ values with those reported in Refs. \cite{Si2025, Dzuba2023}. While Ref. \cite{Si2025} employs a configuration interaction (CI) method, Ref. \cite{Dzuba2023} utilizes a combined approach of linearized RCCSD (SD) and CI (SD+CI) methods. Although all methods yield reasonably consistent results, our present calculations are more precise. The differential FS constants for the 5$F_{7/2}$, 6$D_{3/2}$, and 6$D_{5/2}$ states relative to the ground state are 2.4(19), 32.3(18), and 34.6(18) GHz/fm$^2$, respectively, against the reported values of 2(2), 33(8), and 35(8) GHz/fm$^2$ in Ref. \cite{Berengut2009} that were obtained by accounting for pair-correlation (PC) effects through Br\"uckner orbitals. It is worth noting that our RCC methodology incorporates both PC and core-polarization (CP) effects to all-orders, alongside correlations among PC, CP, non-PC, and non-CP effects together. In the NMS and SMS constants, the most significant corrections originate from Breit interactions, followed by $+$Basis and VP effects.

Utilizing the calculated IS factors, the extracted rms values are summarized in Table \ref{tab-dr}, with the weighted average (denoted by bar) from all considered transitions provided therein. When considering only the FS factors derived from our calculations, we obtain $\overline{\delta\langle r^2\rangle}^{232,229} = 0.305(2)~\mathrm{fm}^2$, which is consistent with the values reported in previous studies: $0.320(15)~\mathrm{fm}^2$ \cite{Dzuba2023} and $0.299(15)~\mathrm{fm}^2$ \cite{Safronova2018-1}. The discrepancies are likely attributable to contributions from triple excitations. Including MS effects, our estimate increases to $0.316(9)~\mathrm{fm}^2$, representing approximately a 3\% enhancement and highlighting the importance of the MS contributions. Additionally, by employing the ratio of isomeric to ISs reported in Ref. \cite{Thielking2018}, we estimate $\overline{\delta\langle r^2\rangle}^{229m,229} = 0.0112(1)~\mathrm{fm}^2$.

\begin{table}[t]
\setlength{\tabcolsep}{4pt}
\centering
\caption{The calculated $B_{hf}/Q_n$ values (in MHz/$b$) for the ground and excited states of $^{229}$Th$^{3+}$, obtained using the same methods as in the previous table, are listed. The weighted average $Q_n$ for $^{229}$Th is then derived by combining these with the measured $B_{hf}$ values. This results in values that agree with some studies but differ substantially from others.}
\begin{tabular}{lrrrr}
\hline\hline
Method & 5$F_{5/2}$ & 5$F_{7/2}$ & 6$D_{3/2}$ & 6$D_{5/2}$ \\[0.3ex]
\hline
DHF & 534.8 & 571.7 & 611.2 & 648.4 \\
RCCSD & 806.1 & 897.1 &  777.1 & 913.5  \\
RCCSDT & 791.9 & 882.1 & 779.3 & 919.2\\[0.3ex]
+Basis & $-1.3$ & $-1.2$ & 2.1 & 2.1 \\[1ex]
+Breit & $-0.7$ & 0.2 & $-3.5$ & $-2.6$ \\[0.3ex]
+VP & $\sim~0.0$ & $\sim~0.0$ & $-0.1$ & 0.1 \\[0.3ex]
\hline
Final & 790(13) & 881(15) & 778(10) & 919(11) \\[0.5ex]
Ref. \cite{Safronova2013} & 725 & 809 & 738 & 873 \\[0.3ex]
Ref. \cite{Berengut2009} & 740 & 860 & 690 & 860 \\[0.3ex]
Ref. \cite{Li2021} & 783 & 879 & 758 & 897 \\[0.3ex]
Ref. \cite{Porsev2021} & 729(10) & 822(13) & 728(6) & 869(11) \\
\hline \\
$B_{hf}$ (Expt.) \cite{Campbell2011} & 2269(2) & 2550(12) & 2265(9) & 2694(7) \\[0.3ex]
$B_{hf}$ (Expt.) \cite{Zitzer2025} & 2270.3(18) &   &  & 2695.7(19) \\
\hline \\
Extracted $Q_n$ & 2.87(5) & 2.89(5) & 2.91(4) & 2.93(4) \\
\hline 
\multicolumn{5}{c}{Recommended $Q_n$ value in $b$} \\
This work & \multicolumn{4}{c}{2.91(3)}\\[0.3ex]
Others  & \multicolumn{4}{c}{3.11(6) \cite{Safronova2013}, 2.80 \cite{Minkov2019}, 3.11(16) \cite{Berengut2009}, } \\[0.3ex]
  &  \multicolumn{4}{c}{2.95(7) \cite{Li2021}, 3.11(2) \cite{Porsev2021} 4.3(9) \cite{Gerstenkorn1974}} \\[0.3ex]
\hline\hline
\end{tabular}
\label{tab3}
\end{table}

It is well established that precise measurements of nuclear moments are crucial for understanding the nuclear structures of isotopes. These values serve as benchmarks for refining nuclear models and enable more reliable predictions of isomeric properties \cite{Zhang2024, Thielking2018, Si2025}. Although several studies have investigated the magnetic dipole (M1) moment ($\mu_I$) and electric quadrupole (E2) moment ($Q_n$) of the $^{229}$Th nucleus, significant discrepancies remain between the predictions of nuclear theory \cite{Minkov2019, Chasman1977} and those derived from combined calculations and measurements of hyperfine structure constants \cite{Campbell2011, Berengut2009, Gerstenkorn1974, Bemis1988, Safronova2013, Li2021, Porsev2021}. In this work, we conduct an independent investigation employing our advanced RCC methods within the expectation value evaluation (EVE) framework.

\begin{figure}[t]
\setlength{\tabcolsep}{3pt}
\centering
\begin{tabular}{cc}
\includegraphics[width=4.0cm,height=4.0cm]{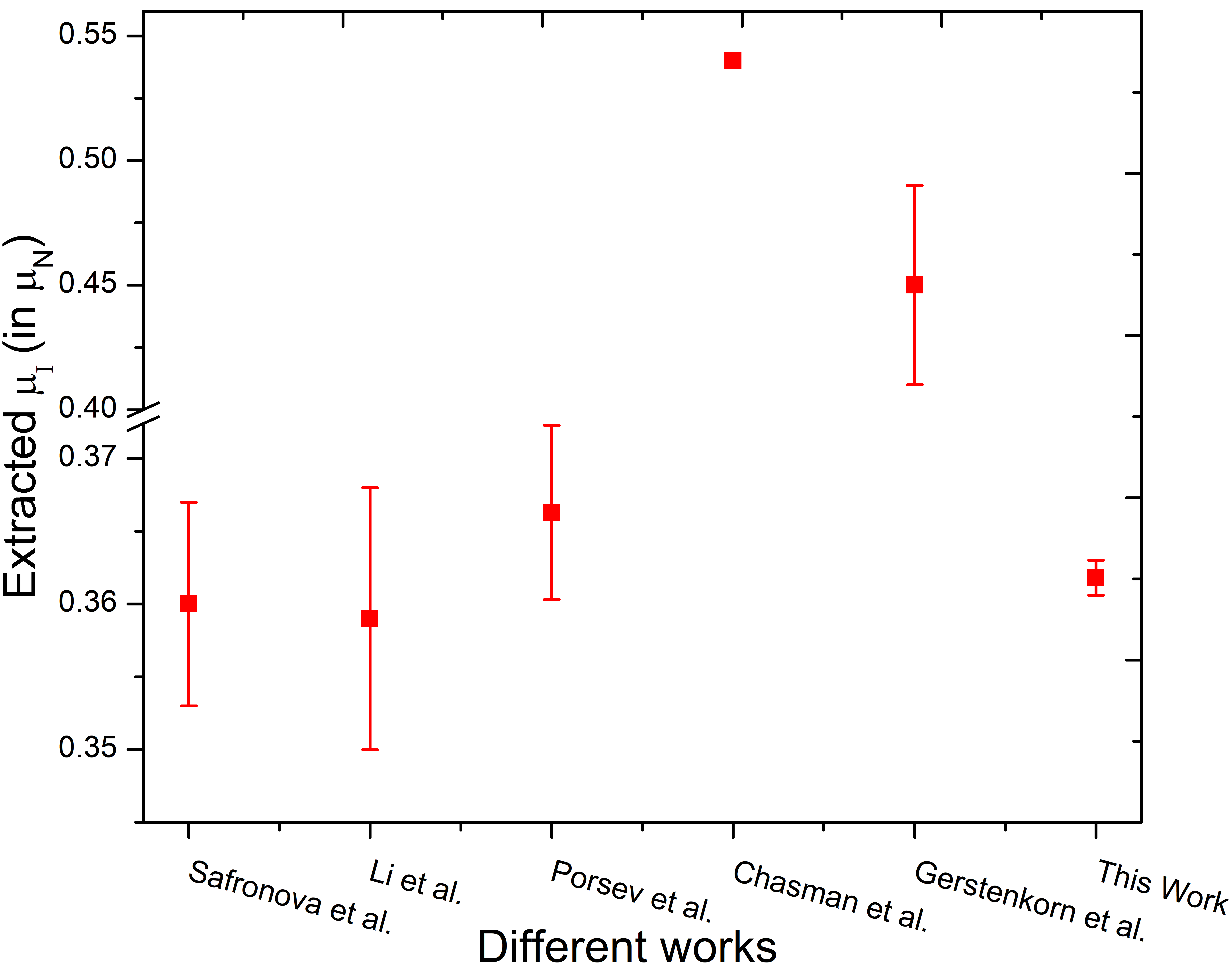}  &  \includegraphics[width=4.0cm,height=4.0cm]{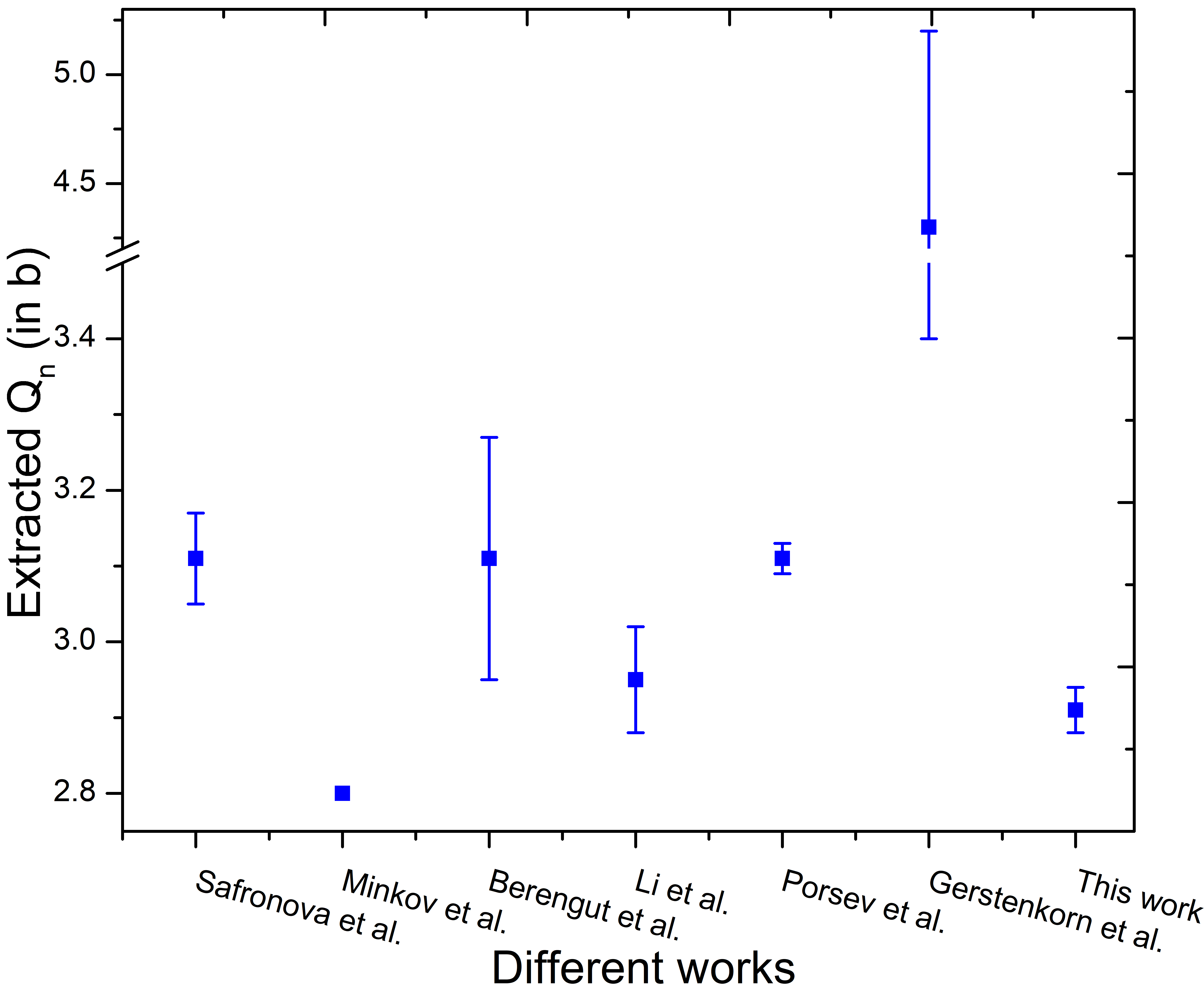}\\
(a) & (b)\\
\end{tabular}
\caption{Comparison of (a) $\mu_I$ and (b) $Q_n$ values of $^{229}\mathrm{Th}$ from different works over the years.}
\label{mu-Q}

\end{figure}

We evaluate the ratios of the M1 hyperfine structure constants, ($A_{hf}$) and $\mu_I$, using the relation $A_{hf}/\mu_I =\langle J||T_{e}^{(1)}||J\rangle/(I\sqrt{J(J+1)(2J+1)}) \mu_N$ where $I$ is the nuclear spin, $J$ is the electronic angular momentum, and $T_{e}^{(1)}$ is the electronic component of the M1 hyperfine interaction as defined in Ref. \cite{charles}. The calculated $A_{hf}/\mu_I$ values for the considered states are obtained using the DHF, RCCSD and RCCSDT methods. These calculations include corrections from $+$Basis, Breit, VP, and Bohr–Weisskopf (BW) effects, as summarized in Table \ref{tab2}. The BW corrections are estimated following a methodology similar to that described in Ref. \cite{bijaya-Yb+} and were found to be negligible. By combining the final $A_{hf}/\mu_I$ ratios with precise experimental $A_{hf}$ values \cite{Campbell2011, Zitzer2025}, we derive $\mu_I$ values from different states, which are listed in the table above. The averaged $\mu_I$ obtained from these results aligns well with previous studies, but with improved precision. Similarly, we evaluate the ratios of the E2 hyperfine constants ($B_{hf}$) and $Q_n$ given by $B_{hf}/Q_n= 2 \sqrt{2J(2J-1)/((2J+1)(2J+2)(2J+3))} \langle J||T_{e}^{(2)}||J\rangle$ where $T_{e}^{(2)}$ represents the electronic component of the E2 hyperfine interaction \cite{charles}. The computed $B_{hf}/Q_n$ values for various states from different theoretical approaches are presented in Table \ref{tab3}. Using measured $B_{hf}$ values from Refs. \cite{Campbell2011, Zitzer2025}, we extract $Q_n$ values for each state, which are also listed in the above table. Notably, the $Q_n$ value derived from this work differs significantly from most previous atomic studies but shows good agreement with nuclear calculations. Figures \ref{mu-Q} (a) and (b) illustrate the predicted $\mu_I$ and $Q_n$ values, respectively, highlighting the contrasting magnitudes reported across different works. Additionally, by employing the ratios of nuclear moments between the isomeric and ground states from Refs. \cite{Zhang2024, Thielking2018}, we estimate $\mu_I = -0.376(54)~\mu_N$ and $Q_n = 1.65(2)$ b for the isomeric state of $^{229}$Th.

Understanding the E2 moments ($\Theta$) of the hyperfine levels, $|FF\rangle$ with $F = I \oplus J$, of the clock states of Th$^{3+}$ is essential for accurately estimating E2 shifts in the clock transitions depicted in Fig. \ref{fig-clock}. Following Ref. \cite{Derevianko}, the E2 moment ($\Theta(F)$) of the $|FF\rangle$ level can be expressed as a perturbative expansion: $\Theta(F)= \Theta(F)^{(0)} + \Theta(F)^{(1)}$, where the superscripts denote the order with respect to the hyperfine interaction. The $\Theta(F)^{(0)}$ relates directly to the atomic E2 moment $\Theta(J) = \langle J, J | Q^{(2)} | J, J \rangle$, with the E2 operator $Q^{(2)} = \frac{1}{2} \sum_i \left( 3 z_i^2 - r_i^2 \right)$, given by
\begin{eqnarray}
\Theta(F)^{(0)}= \Theta(J) (-1)^{I+J+F+2} (2F+1) \left( \begin{array}{ccc}
                J & 2 & J \\
                -J & 0 & J 
\end{array}\right)^{-1} \nonumber \\ \times \left( \begin{array}{ccc}
                F & 2 & F\\
                -F & 0 & F 
\end{array}\right)  \left\{ \begin{array}{ccc}
                J & F & I \\
                F & J & 2 
\end{array}\right\} . \ \ \ \ \ \ \ \ \ \ \ 
\end{eqnarray}
The $\Theta(F)^{(1)}$ part is expressed as 
\begin{eqnarray}
\Theta(F)^{(1)} = 2(2F+1) (-1)^{2F+2I} \left( \begin{array}{ccc}
                F & 2 & F\\
                -F & 0 & F 
\end{array}\right) \ \ \ \ \ \ \ \ \nonumber\\
\times \sum_{k,J'} (-1)^{J+J'} \left\{ \begin{array}{ccc}
                I & I & k\\
                J & J' & F 
\end{array}\right\}\left\{ \begin{array}{ccc}
                F & 2 & F\\
                J' & I & J 
\end{array}\right\} \ \ \ \nonumber\\
\times \langle I||T^{(k)}_{n}||I\rangle \frac{\langle J||Q^{(2)}||J'\rangle\langle J'||T^{(k)}_{e}||J\rangle}{E_J-E_{J'}} , \ \ \ \ \ \ \ 
\end{eqnarray}
where $E_J$s denotes atomic energies, $k$ is summed over the M1 ($k=1$) and E2 ($k=2$) components, respectively, $\langle I||T^{(1)}_{n}||I\rangle= \mu_I \mu_N \sqrt{(I+1)(2I+1)/I}$ and $\langle I||T^{(2)}_{n}||I\rangle = Q_n \sqrt{(2I-2)! (2I+3)!}/(2(2I)!)$. In Table \ref{tab-E2}, we present $\Theta(J)$ values (in atomic units, a.u.) for the $5F_{5/2}$ state obtained through various approximations, along with comparisons to previously reported values \cite{Yu2024, Safronova2013-pol}, highlighting notable differences. The $\Theta(F)^{(0)}$ value for the $F=5$ level of the ground state and the $F=4$ level of the isomeric state are found to be identical, both being 0.614(9) a.u.. However, the $\Theta(F)^{(1)}$ values for the $F=5$ level of the ground state are $-0.44 \times 10^{-6}$ a.u. for $k=1$ and $-2.09 \times 10^{-6}$ a.u. for $k=2$. Similarly, for the isomeric $F=4$ level involved in the clock transition, the $\Theta_F^{(1)}$ values are  $0.46 \times 10^{-6}$ a.u. and $-1.19 \times 10^{-6}$ a.u. for $k=1$ and $k=2$, respectively.

\begin{table}
\setlength{\tabcolsep}{10pt}
\centering
\caption{Our {\it ab initio} calculated ground state $\Theta$, $\alpha_d^S$ and $\alpha_d^T$ values (in a.u.) of Th$^{3+}$ along with a comparison to reported values from earlier studies.}
\begin{tabular}{lrrr}
\hline\hline
Method & $\Theta$ & $\alpha^s$ & $\alpha^t$ \\[0.3ex]
\hline
DHF &  0.9160 & $-28.53$ &  15.32 \\[0.3ex]
RCCSD & 0.6266 & 15.08  & $-5.53$ \\[0.3ex]
RCCSDT  & 0.6188 & 14.99 & $-5.44$ \\[0.3ex]
+Basis & $-0.0016$ & $-0.07$ & $0.07$  \\[1ex]
+Breit & $-0.0039$ & $-0.59$ & 0.59  \\[0.3ex]
+VP & 0.0003 & $\sim~0.0$ & $\sim~0.0$   \\[0.3ex]
Final  & 0.614(9) & 14.33(26) & $-4.78(20)$ \\[0.3ex]
Ref. \cite{Campbell2012} &  & 13 & $-4.2$ \\
Ref. \cite{Yu2024} & 0.6237 & 14.16 & $-6.15$ \\
Ref. \cite{Safronova2013-pol}&  0.624(14) & 14.67(60) & $-6.07(53)$ \\ 
\hline\hline
\end{tabular}
\label{tab-E2}
\end{table}

Accurate determination of the electric dipole (E1) polarizability, $\alpha_d$, is essential for precise estimation of Stark and black-body radiation shifts in the clock transition. Previously, both the scalar ($\alpha_d^S$) and tensor ($\alpha_d^T$) components of $\alpha_d$ for the ground state of Th$^{3+}$ were evaluated using the sum-over-states (SOS) approach. However, the SOS method inherently faces limitations in uniformly accounting for contributions from all intermediate states, often requiring a combination of approaches—particularly to incorporate effects from doubly excited and higher-lying configurations. To treat all intermediate states on an equal footing, we determine $\alpha_d^{S}$ and $\alpha_d^{T}$ through an {\it ab initio} methodology based on the expressions  $\alpha_d^{S/T} = \langle \Psi_v^{(0)}|\tilde{D}^{S/T}|\Psi_v^{(1)} \rangle$, where $\tilde{D}^{S/T}$ are the effective E1 operators as defined in Refs. \cite{Chakraborty2025-Li, Chakraborty2025-Cs}, and $|\Psi_v^{(1)} \rangle$ is the first-order perturbed wave function obtained by solving the inhomogeneous equation
\begin{eqnarray}
(H_{at} - E_v^{(0)}) |\Psi_v^{(1)} \rangle = - D |\Psi_v^{(0)} \rangle,
\end{eqnarray}
with the E1 operator $D$.

The {\it ab initio} values of $\alpha_d^S$ and $\alpha_d^T$ computed at various levels of approximation are summarized in Table \ref{tab-E2}. These results are compared with previously reported SOS semi-empirical values. As evident from the table, our findings are in good agreement with existing data, but they are obtained with considerably higher precision. Although the differences between the RCCSD and RCCSDT results for both quantities are relatively small, we emphasize the importance of including contributions from orbitals with $l>6$ to achieve the desired accuracy. For example, the RCCSD result for $\alpha_d^S$, when incorporating orbitals up to $g$-symmetry, is approximately 20.0 a.u., which is about 30\% higher than the value reported in the table obtained using the RCCSD method.

BKS is supported by ANRF with grant no. CRG/2023/002558 and Department of Space, Government of India. All calculations were performed on the ParamVikram-1000 HPC cluster at Physical Research Laboratory (PRL), Ahmedabad, India.


\begin{thebibliography}{50}

\bibitem{Kraemer2023}
S. Kraemer, J. Moens, M. Athanasakis-Kaklamanakis, S. Bara, K. Beeks, P. Chhetri, K. Chrysalidis, A. Claessens, T. E. Cocolios, J. G. M. Correia et al., Nature (London) {\bf 617}, 706 (2023).

\bibitem{Peik2003}
E. Peik and C. Tamm, Europhys. Lett. {\bf 61}, 181 (2003).

\bibitem{Peik2021}
E. Peik, T. Schumm, M. S. Safronova, A. Pálffy, J. Weitenberg, and P. G. Thirolf, Quantum Sci. Technol. {\bf 6}, 034002 (2021).

\bibitem{Campbell2012}
C. J. Campbell, A. G. Radnaev, A. Kuzmich, V. A. Dzuba, V. V. Flambaum, and A. Derevianko, Phys. Rev. Lett. {\bf 108}, 120802 (2012).

\bibitem{Flambaum2006}
V. V. Flambaum, Phys. Rev. Lett. {\bf 97}, 092502 (2006).

\bibitem{Safronova2018}
M. S. Safronova, D. Budker, D. DeMille, D. F. J. Kimball, A. Derevianko, and C. W. Clark, Rev. Mod. Phys. {\bf 90}, 025008 (2018).

\bibitem{Campbell2011}
C. J. Campbell, A. G. Radnaev, and A. Kuzmich, Phys. Rev. Lett. {\bf 106}, 223001 (2011).

\bibitem{Zitzer2025}
G. Zitzer, J. Tiedau, Ch. E. Düllmann, M. V. Okhapkin, and E. Peik, Phys. Rev. A {\bf 111}, L050802 (2025).

\bibitem{Safronova2013}
M. S. Safronova, U. I. Safronova, A. G. Radnaev, C. J. Campbell, and A. Kuzmich, Phys. Rev. A {\bf 88}, 060501(R) (2013).

\bibitem{Berengut2009}
J. C. Berengut, V. A. Dzuba, V. V. Flambaum, and S. G. Porsev, Phys. Rev. Lett. {\bf 102}, 210801 (2009).

\bibitem{Dzuba2023}
V. A. Dzuba and V. V. Flambaum, Phys. Rev. Lett. {\bf 131}, 263002 (2023).

\bibitem{Si2025}
Ran Si, Chaofan Shi, Nan Xue, Xiangjin Kong, Chongyang Chen, Bingsheng Tu, Yu-Gang Ma, Sci. China Phys. Mech. Astron. {\bf 68}, 272011 (2025).

\bibitem{Safronova2018-1}
M.S. Safronova, S. G. Porsev, M.G. Kozlov, J. Thielking, M.V. Okhapkin, P. Glowacki, D.M. Meier, and E. Peik, Phys. Rev. Lett. {\bf 121}, 213001 (2018).

\bibitem{King}
W. H. King, {\it Isotope shifts in atomic spectra}, Springer Science \& Business Media, (2013).

\bibitem{Sahoo-review}
B. K. Sahoo, S. Blundell, A. V. Oleynichenko, R. F. G. Ruiz, L. V. Skripnikov, and B. Ohayon, J. Phys. B {\bf 58}, 042001 (2025).

\bibitem{Chakraborty2025-Li}
A. Chakraborty and B. K. Sahoo, Phys. Rev. A {\bf 111}, 042807 (2025). 

\bibitem{Chakraborty2025-Cs}
A. Chakraborty and B. K. Sahoo, Phys. Rev. A {\bf 111}, 062812 (2025). 

\bibitem{Katyal2025}
Vaibhav Katyal, A. Chakraborty, B. K. Sahoo, Ben Ohayon, Chien-Yeah Seng, Mikhail Gorchtein, and John Behr, Phys. Rev. A {\bf 111}, 042813 (2025).

\bibitem{Porsev2021}
S. G. Porsev, M. S. Safronova, and M. G. Kozlov, Phys. Rev. Lett. {\bf 127}, 253001 (2021).

\bibitem{NIST}
A. Kramida, Yu. Ralchenko, J. Reader, and NIST ASD Team, NIST Atomic Spectra Database, version 5.8, https://physics.nist.gov/asd (2020).

\bibitem{Blaise} 
J. Blaise and J. Wyart, {\it Selected constants, energy levels, and atomic spectra of actinides}, https://www.lac.universite-paris-saclay.fr/Data/Database/.

\bibitem{Thielking2018}
J. Thielking, M. V. Okhapkin, P. Glowacki, D. M. Meier, L. von der Wense, B. Seiferle, C. E. Dullmann, P. G. Thirolf, and E. Peik, Nature {\bf 556}, 321 (2018).

\bibitem{Zhang2024}
C. Zhang, Tian Ooi, Jacob S. Higgins, Jack F. Doyle, Lars von der Wense, Kjeld Beeks, Adrian Leitner, Georgy A. Kazakov, Peng Li, Peter G. Thirolf, Thorsten Schumm, and Jun Ye, Nature {\bf 633}, 63 (2024).

\bibitem{Chasman1977}
R. R. Chasman, I. Ahmad, A. M. Friedman, and J. R. Erskine, Rev. Mod. Phys. {\bf 49}, 833 (1977). 

\bibitem{Minkov2019}
N. Minkov and A. Pálffy, Phys. Rev. Lett. {\bf 122}, 162502 (2019).

\bibitem{Gerstenkorn1974}
S. Gerstenkorn, P. Luc, J. Verges, D. W. Englekemeir, J. E. Gindler, and F. S. Tomkins, J. Phys. (Paris) 35, 483 (1974).

\bibitem{Bemis1988}
C. E. Bemis, F. K. McGowan, J. L. C. Ford, Jr., W. T. Milner, R. L. Robinson, P. H. Stelson, G. A. Leander, and C. W. Reich, Phys. Scr. {\bf 38}, 657 (1988).

\bibitem{Li2021}
F.-C. Li, H.-X. Qiao, Y.-B. Tang, and T.-Y. Shi, Phys. Rev. A {\bf 104}, 062808 (2021).

\bibitem{charles}
C. Schwartz, Phys. Rev. {\bf 97}, 380 (1955).

\bibitem{bijaya-Yb+}
B. K. Sahoo, Phys. Rev. A {\bf 111}, L060801 (2025).

\bibitem{Derevianko}
A. Derevianko, Phys. Rev. A {\bf 93}, 012503 (2016).
 
\bibitem{Yu2024}
Shi-Cheng Yu, Wen-Ting Gan, Xia Hua, Xin Tong and Cheng-Bin Li, Phys. Rev. A {\bf 109}, 063115 (2024).

\bibitem{Safronova2013-pol}
M. S. Safronova and U. I. Safronova, Phys. Rev.A {\bf 87}, 062509 (2013).

\end{thebibliography}
\end{document}